# Magnetic and Magnetocaloric Study of the Ferromagnetically Coupled GdF$_3$: The Best Cryogenic Magnetic Coolant Ever

Yan-Cong Chen,[a] Jun-Liang Liu[a] and Ming-Liang Tong*[a]

The magnetic susceptibility and isothermal magnetization for GdF$_3$ were measured, and the isothermal entropy change was evaluated up to 9 T. Combining the large isotropic spin of Gd$^{3+}$, the dense structure and the weak ferromagnetic interaction, an extremely large -$\Delta S_m$ for GdF$_3$ was observed up to 528 mJ cm$^{-3}$ K$^{-1}$ for $\Delta H$ = 9 T, proving itself to be the best cryogenic magnetic coolant ever.

## Introduction

The magnetocaloric effect (MCE) was discovered in 1881 in metallic iron by Warburg,[1] and the magnetic refrigeration soon become a powerful technique to obtain and maintain ultra-low temperature by adiabatic demagnetization (ADR).[2-3] At an early stage, the cryogenic magnetic coolants are mainly the inorganic paramagnetic salts and oxides, such as Gd$_2$(SO$_4$)$_3$·8H$_2$O and Gd$_3$Ga$_5$O$_{12}$ (GGG).[4] In recent years, molecular materials suddenly emerged in this field as an unprecedented classification, and a lot of highly competitive molecular magnetic coolants have been synthesized and characterized.[5-7] Their distinct advantages, such as stoichiometric composition, monodispersity and modificability, have provided the researchers a perfect platform to realize the design strategies towards cryogenic magnetic coolants. Various lessons have been learned and proved, including but not limited to the large spin state, low anisotropy, weak interaction and large metal-to-ligand ratio.[8]

For the best cooling performance around liquid helium temperature, it is believed that the Gd$^{3+}$ ion is a wonderful choice owing to the half-filled 4f orbital ($S$ = 7/2), magnetic isotropy and usually weak intermetallic interactions. Additionally, the low magnetic ordering temperature can still be maintained even with high metal-to-ligand ratio, and a large MCE can be obtained, especially with volumetric units.[9] Soon after, the competition in molecular magnetic coolants became the race on the reduction of counterions accompanying Gd$^{3+}$. To date, many gadolinium carboxylates including acetates, formats and oxalates have been reported, with increasing MCE chasing after GGG.[10]

Since there is little room for the organic ligands to keep shrinking, the most recent focus in this field dramatically returned to the inorganic compounds based on small counterions such as OH$^-$, SO$_4^{2-}$, O$^{2-}$.[11] Following the strategies learned in molecular systems, significant increase of MCE in these inorganic compounds have been observed, and the orthorhombic Gd(OH)CO$_3$ finally surpass the performance of GGG and set a new record.[12] However, the story seems not over yet as a better example of GdPO$_4$ was reported just after half a year.[13]

In the inorganic area, the utilization of the strategy must be more careful, because strong magnetic interactions are more likely to exist, thus hinder the MCE. Through intensive literature research, we are excited to found that the gadolinium fluoride, GdF$_3$, have been tested for a toroidal ADR prototype.[14] Despite some Gd$_2$O$_3$ impurity, the results already showed superior performance than GdPO$_4$ for $\Delta H$ = 5 T. Therefore, here we present the detailed magnetic and magnetocaloric study on GdF$_3$, reveals the ferromagnetic coupling and extremely large MCE, proving it to be the best cryogenic magnetic coolant ever.

## Results and Discussion

Pure polycrystalline GdF$_3$ can be hydrothermally synthesized from aqueous GdCl$_3$ and excess amount of NaBF$_4$ or NH$_4$F, while commercial product is also available. The Experimental Powder X-ray diffraction was measured on polycrystalline sample at room temperature on a Bruker D8 X-Ray Diffractometer with Cu K$\alpha$ radiation, which is in line with the reference JCPDF #49-1804 (**Figure 1**). Pattern indexing and cell refinement were performed with MDI Jade software, giving the the consistent cell parameters with the reference (**Table 1**).

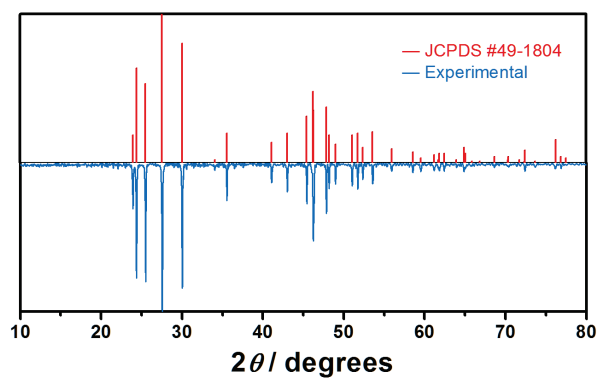

**Figure 1** Experimental powder XRD pattern of GdF$_3$ compared to the reference (JCPDF #49-1804)

GdF$_3$ crystallizes in the orthorhombic space group *Pnma* and the same structure for SmF$_3$, EuF$_3$, TbF$_3$, HoF$_3$ and YbF$_3$ were also reported. In the crystal structure, each Gd$^{3+}$ is 9-coordinated and each F$^-$ is $\mu_3$-bridging, just corresponding to the stoichiometric ratio 1:3. Thanks to the simple composition, the formula mass of GdF$_3$ is only 214.25 g mol$^{-1}$, trapping as much as 73.4% of Gd$^{3+}$ in an extremely dense structure with $\rho$ = 7.063 g cm$^{-3}$. Therefore, GdF$_3$ shall be a wonderful magnetic coolant so long as there isn't long-range antiferromagnetic ordering in the working temperature region.

**Table 1** Crystal Cell Refinement for GdF₃ from the experimental powder XRD pattern compared to the reference (JCPDF #49-1804).

|  | Experimental | JSPDF #49-1804 |
|---|---|---|
| Radiation | Cu Kα | Cu Kα |
| Crystal system | Orthorhombic | Orthorhombic |
| Space group | *Pnma* | *Pnma* |
| Z | 4 | 4 |
| $a$/Å | 6.5715(6) | 6.571 |
| $b$/Å | 6.9829(6) | 6.984 |
| $c$/Å | 4.3903(5) | 4.390 |
| Unit cell volume/Å³ | 201.46 | 201.47 |
| $\rho_{calcd}$ / g cm⁻³ | 7.0635 | 7.063 |

Variable-temperature magnetic susceptibility measurement was performed on polycrystalline sample of GdF₃ in an applied dc field of 0.1 T (**Figure 2**), using a Quantum Design PPMS with VSM option. At room temperature, the $\chi_m T$ value is 7.96 cm³ K mol⁻¹, slightly larger than the spin-only value expected for a free Gd³⁺ ion with $g$ = 2 (7.875 cm³ K mol⁻¹). Upon cooling, $\chi_m T$ keeps increasing to 11.2 cm³ K mol⁻¹ at 2 K, suggesting dominant ferromagnetic interactions between Gd³⁺ ions. The inverse magnetic susceptibility ($1/\chi_m$) obeys the Curie-Weiss law with $C$ = 7.91 cm³ K mol⁻¹ and $\theta$ = +0.7 K. No sign of the long-range magnetic ordering can be observed above 2 K, as the ordering temperature is reported to be around 1.25 K.[15] Such a behaviour is extremely favourable for a large cryogenic MCE, and it is quite different from those complexes with hydroxide bridges, where strong antiferromagnetic interactions are common and usually harm the full utilization of the MCE. The ferromagnetic coupling in GdF₃ and the low ordering temperature rule out our last worry about its capability as a promising cryogenic magnetic coolant.

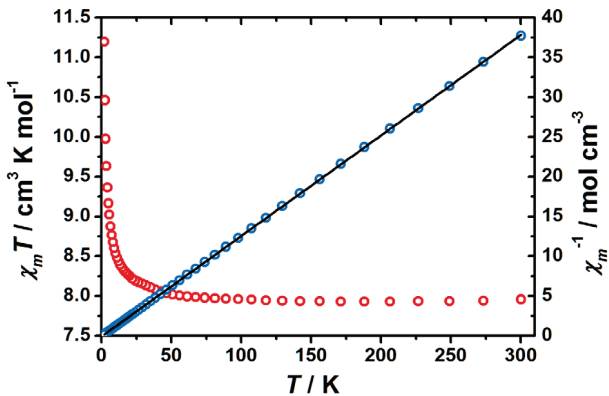

**Figure 2** Temperature-dependencies of the magnetic susceptibility product ($\chi_m T$) and inverse magnetic susceptibility ($1/\chi_m$) at 2-300 K with a dc field of 0.1 T. The black solid line represents the least-square fit for the Curie-Weiss law.

To calculate the precise value of $-\Delta S_m$, the isothermal magnetization for GdF₃ were measured from 2 K to 10 K in an applied dc field up to of 9 T (**Figure 3**). The magnetization increases quickly with the applied field below 2 T and reaches the saturation value of 7.0 $N\beta$ at 2 K and 9 T, which is in good agreement with the expected value for a Gd³⁺ ion ($s = 7/2$, $g$ = 2).

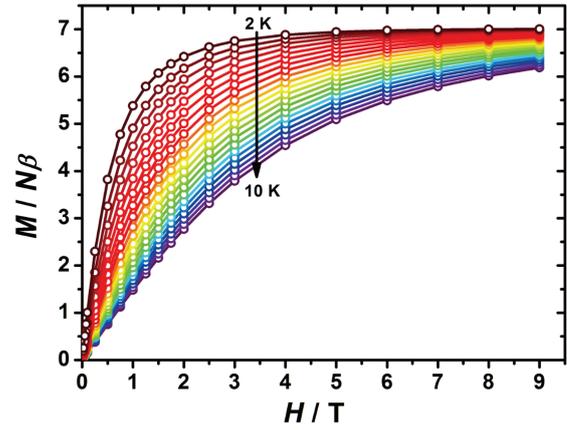

**Figure 3** Magnetization *versus* the dc field in the temperature range of 2-10 K.

The isothermal entropy change can be calculated by applying the Maxwell equation:

$$\Delta S_m(T) = \int_0^H \left[\partial M(T, H)/\partial T\right]_H dH$$

Just as many reported ferromagnetic coupling systems, the maximum $-\Delta S_m$ values for GdF₃ grow rapidly with increasing $\Delta H$, namely 181 mJ cm⁻³ K⁻¹, 321 mJ cm⁻³ K⁻¹ and 399 mJ cm⁻³ K⁻¹ for $\Delta H$ = 1 T, 2 T and 3 T, respectively. For larger $\Delta H$, the increase of $-\Delta S_m$ values become slower, reaching 474 mJ cm⁻³ K⁻¹ and 506 mJ cm⁻³ K⁻¹ for $\Delta H$ = 5 T and 7 T, and the peaks of $-\Delta S_m$ *versus* $T$ curves gradually shift to higher temperatures. These results are in line with the heat capacity tests for the toroidal ADR prototype,[14] and the maximum value here we obtained is 528 mJ cm⁻³ K⁻¹ (74.8 J kg⁻¹ K⁻¹ )at $T$ = 3.2 K and $\Delta H$ = 9 T, close to the theoretical limiting value of 570 mJ cm⁻³ K⁻¹ (80.7 J kg⁻¹ K⁻¹) calculated from $R\ln(2s+1)/Mw$ with $s = 7/2$ and $Mw$ = 214.25 g mol⁻¹.

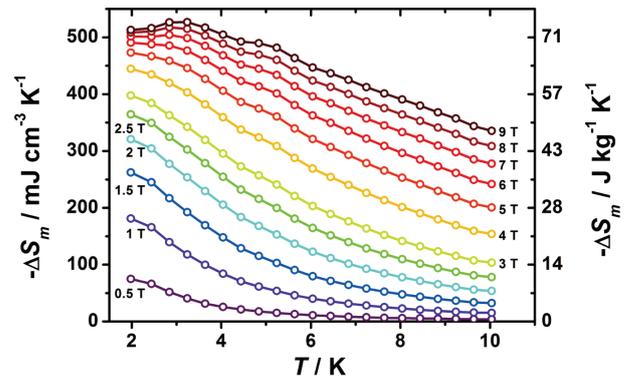

**Figure 4** Temperature-dependencies of $-\Delta S_m$ for selected $\Delta H$ obtained from magnetization. The data with field variation below 0.5 T are omitted for clarity.

## Concluding Remarks

The area of cryogenic magnetic coolants is full of competitions rather than a monopolized industry, and there has never been a complex that can hold the record for a long time. Thanks to the competing peers in a sense, researchers now are much wiser at adopting suitable strategies for the design of high performance magnetic coolants. As the record set by Gd(OH)CO₃[12] has been taken by the recently reported GdPO₄,[13] we have to fight with our back to the river and present the ferromagnetically coupled GdF₃.

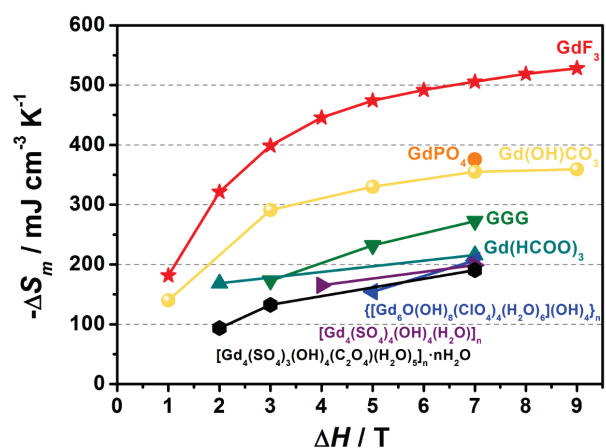

**Figure 5** The maximum reported -$\Delta S_m$ value *versus* the corresponding $\Delta H$ for selected cryogenic magnetic coolants.[4, 10d-e, 11-13]

As demonstrated in **Figure 5**, the magnetic coolant with -$\Delta S_m$ larger than 400 mJ cm$^{-3}$ K$^{-1}$ has never been reported, while the maximum value for GdF$_3$ has exceeded 500 mJ cm$^{-3}$ K$^{-1}$. Comparing the -$\Delta S_m$ with $\Delta H$ = 7 T for the sake of fairness, the performance of GdF$_3$ (506 mJ cm$^{-3}$ K$^{-1}$) surpass that of GdPO$_4$ (375.8 mJ cm$^{-3}$ K$^{-1}$) by 34.6%, setting a new record. Last but not least, the dense structure and ferromagnetic interaction in GdF$_3$ leads to still excellent performance even for lower $\Delta H$ such as 2 T and 1 T, highlighting the competitiveness.

At this point, we believe the long journey in pursuit of the best cooling performance is close to the extreme. There is limited room, if any, for the further increase of the -$\Delta S_m$ value itself for Gd-based materials: some other compounds like GdOF, GdOOH, GdBO$_3$ and Gd$_2$(CO$_3$)$_3$ might worth a try, while Gd$_2$O$_3$ and Gd(OH)$_3$ are already known as antiferromagnets. Future study on the cryogenic MCE shall turn the focus onto the other parameters such as $\Delta T_{ad}$, cooling power and even production cost, where the Mn-based materials become strong competitors.[5g] Although we have to admit that the coordination complexes with organic ligands can never be so comparative with inorganic complexes, we also should not forget that the ambitions of molecular materials have never been just about the value. We have witnessed how the numerous complexes seems useless can be rationally modified into good candidates, and we have been continuously learning about the magneto-structural correlations during the the design and synthesis of these molecular magnetic coolants. Furthermore, chemists' powerful skill on material engineering can still modify the behaviour of the inorganic compounds in the limitless nanomaterial and organic-inorganic hybrid material, where the magnetic interaction, ordering temperature and low-field performance may be optimized. Finally, we would like to quote what Churchill stated during WWII:

*"This is not the end. It is not even the beginning of the end. But it is, perhaps, the end of the beginning."*

### Acknowledgements
This work was supported by the "973 Project" (2012CB821704 and 2014CB845602), project NSFC (Grant no. 91122032, 21371183, 21121061 and 21201137), the NSF of Guangdong (S2013020013002), Program for Changjiang Scholars and Inno-vative Research Team in University of China (IRT1298)

### Notes and references
[a] Key Laboratory of Bioinorganic and Synthetic Chemistry of Ministry of Education, State Key Laboratory of Optoelectronic Materials and Technologies, School of Chemistry and Chemical Engineering, Sun Yat-Sen University, Guangzhou 510275, P. R. China
E-mail: tongml@mail.sysu.edu.cn (M.-L. Tong)